\documentclass[a4paper,12pt]{article}

\usepackage{amsmath,amssymb}
\usepackage{graphicx}
\usepackage{lmodern}
\usepackage[breaklinks,bookmarks=true,hyperfigures=true]{hyperref}
\usepackage{subfig}
\usepackage{tikz}
\usepackage{rotating}
\usepackage{pgf,pgfarrows,pgfnodes}
\usepackage{longtable}
\usepackage{pdfpages}
\usepackage[a4paper,text={160mm,225mm},centering]{geometry}

\newcommand{\op}{{\mathcal{O}}}
\newcommand{\nn}{\nonumber}
\newcommand{\tr}{{\rm Tr\,}}

\newcommand{\sym}{{\mathcal{N}=4~ \text{SYM}}}
\newcommand{\syml}{\mathcal{N}=4~ \text{super Yang-Mills}}

\begin{document}

\thispagestyle{empty}

\null\vskip-47pt \hfill
\begin{minipage}[t]{50mm}
~~~~~~~~~~~~~~~~~HU-EP-12/23\\
\end{minipage}

\vskip2.2truecm
\begin{center}
\vskip 0.2truecm

 {\Large\bf
Three-Point Functions of Twist-Two Operators \\[.3cm] in $\mathcal{N}=4$ SYM at One Loop}
\vskip 0.5truecm

\vskip 1truecm
{\bf    Jan Plefka$^{a}$, Konstantin Wiegandt$^{a,b}$\\
}

\vskip 0.4truecm
$^{a}$ {\it Institut f\"ur Physik, Humboldt-Universit\"at zu Berlin,
\\
Newtonstra\ss e 15, 12489 Berlin}
 \\
  \vskip .2truecm
$^{b}$ {\it  II. Institut f\"ur Theoretische Physik, Universit\"at Hamburg,\\ Luruper Chaussee 149, 22716 Hamburg}
\end{center}

~\\[1.5cm]

\centerline{\bf Abstract} 
\medskip
\noindent
We calculate three-point functions of two protected operators and one twist-two operator with arbitrary even spin $j$ in $\syml$ theory to one-loop order. In order to carry out the calculations we project the indices of the spin $j$ operator to the light-cone and evaluate the correlator in a soft-limit where the momentum coming in at the spin $j$ operator becomes zero. This limit largely simplifies the perturbative calculation, since all three-point diagrams effectively reduce to two-point diagrams and the dependence on the one-loop mixing matrix drops out completely. 
The results of our direct calculation are in agreement with the structure constants obtained by F.A. Dolan and H. Osborn from the operator product expansion of four-point functions of half-BPS operators.

\newpage

\section{Introduction and Conclusions}
The full solution of $\sym$ theory is equivalent to the ability of calculating any observable of interest in this quantum field theory exactly or perturbatively to arbitrary order in the coupling constant. In recent years, such impressive achievements have indeed been made for certain observables in planar $\sym$, largely due to the existence of integrability \cite{Minahan:2002ve, Beisert:2003tq, Bena:2003wd,Beisert:2003yb} in the AdS/CFT correspondence
\cite{Maldacena:1998re,Witten:1998qj,Gubser:1998bc}, see \cite{Beisert:2010jr} for a review.

Using the methods of integrability, all-order conjectures, e.g. for the cusp anomalous dimension of twist two operators could be obtained \cite{Staudacher:2004tk,Beisert:2006ez}. Using this, also for scattering amplitudes of four and five external
particles  in planar $\sym$, all-order results became available \cite{Bern:2005iz,Drummond:2008vq}. These all-order results were achieved using the duality between MHV amplitudes and light-like Wilson loops \cite{Alday:2007hr,Drummond:2007aua,Brandhuber:2007yx}, see \cite{Alday:2008yw,Henn:2009bd} for reviews.

Given these results, it is natural to ask, whether similar advances can be made for three-point functions of gauge invariant operators. Not only are three-point functions the next natural correlators to address after two-point functions, but they possess a particular importance due to the operator product expansion (OPE). Given the knowledge of all structure constants that appear in the theory, one can in principle construct any higher-point correlation function using the OPE. Therefore, all-order results for structure constants, together with all-order results for anomalous dimensions yield important information on any higher-point function.

The form of three-point functions is fixed by conformal symmetry and the only dependence on the coupling constant is contained in the anomalous dimensions of the operators and the structure constants
which receive radiative corrections in the coupling 
\begin{equation}
C_{\alpha\beta\gamma}(g^2) = C_{\alpha\beta\gamma}^{(0)} + g^2 C_{\alpha\beta\gamma}^{(1)} + ...
\end{equation}
Similarly as for two-point functions, there are non-renormalisation theorems for three-point correlators of half-BPS operators, which guarantee that they do not get quantum corrections \cite{Lee:1998bxa,Eden:1999gh,Arutyunov:2001qw,Heslop:2001gp,Basu:2004nt,Baggio:2012rr}. 

Direct computations of three-point functions in $\sym$ theory have been performed in \cite{Bianchi:2001cm,Beisert:2002bb,Chu:2002pd,Roiban:2004va,Okuyama:2004bd,Alday:2005nd,Alday:2005kq,Georgiou:2008vk,Georgiou:2009tp}. In \cite{Georgiou:2012zj} a large number of three-point functions involving scalar primary operators of up to and including length five is considered.

The role of integrability for three-point function calculations was first addressed in \cite{Okuyama:2004bd,Roiban:2004va,Alday:2005nd}. Applications of integrability methods  can be found in \cite{Escobedo:2010xs,Escobedo:2011xw,Gromov:2011jh,Kostov:2012jr} at tree-level and in \cite{Gromov:2012vu,Kostov:2012jr} for loop-level three-point functions of scalar single trace operators. It does indeed turn out, that three-point functions can be studied efficiently using integrability.

The question of a weak-strong coupling matching was addressed in \cite{Escobedo:2011xw,Bissi:2011dc,Bissi:2011ha,Grignani:2012yu,Grignani:2012ur,Georgiou:2011qk}. The SL(2) sector was adressed in \cite{Georgiou:2011qk} and agreement between the structure constants at weak and strong coupling was found, where a BPS state and two operators with large spin were considered at tree-level.  It would be interesting to promote this study to the loop-level, which requires the knowledge of the correlator of one BPS operator and two operators with spin at one-loop level. Furthermore, it would be interesting to see, how one can make use of integrability methods in order to calculate this correlator at loop-level.

Here, we make a first step towards this direction by providing a direct field theory computation of three-point correlators with two protected operators and one twist-two operator. This calculation serves as a starting point for the case of two operators with spin, where several distinct space-time structures enter with a priori different structure constants, see e.g.  \cite{Fradkin:1996is},  \cite{Sotkov:1976xe}. A direct calculation is possibly more suitable for the development of integrability methods than the extraction of the same coefficient from the OPE.\\

More explicitly, here we calculate three-point functions in $\sym$ theory
 involving two protected scalar operators
\begin{align}\label{eqn:BPS-operators}
\op(x) = \tr \left( \bar{\phi}_{12}(x)  \bar{\phi}_{13}(x) \right),\qquad
 \tilde{\op}(x) = \tr \left( \bar{\phi}_{12} (x) \phi^{13}(x) \right)
\end{align}
and one twist-two operator with arbitrary even spin $j$, which is totally symmetric and traceless in all indices and schematically of the form 
\begin{equation}
\op_{\mu_1..\mu_j} (x) = \tr \left( D_{\mu_1}..D_{\mu_k} \phi^{12}(x)D_{\mu_{k+1}}..D_{\mu_{j}} \phi^{12}(x)\right)+...
\end{equation}
where the ellipses stand for a certain distribution of the derivatives, which is given explicitly in section \ref{sec:calculation}.
Conformal symmetry fixes the three-point functions up to the structure constants $C_{\op\tilde{\op}j}$. We use the light-cone projection, see appendix \ref{sec:light-cone-projection}, where all indices are contracted with a light-like vector $z^\mu$.  Then the correlator with the renormalised operator $\hat{\mathbb{O}}_j$, see  \eqref{equ:renormalised-op}, reads 
\begin{equation}\label{eqn:three-point-structure}
 \langle \op(x_1)  \tilde{\op}(x_2) \hat {\mathbb{O}}_j(x_3) \rangle = C_{\op \tilde{\op} j}(g) \frac{(  \hat{Y}_{12,3}) ^j}{|x_{12}|^{\Delta_1 + \Delta_2 - \theta} |x_{13}|^{\Delta_1 +  \theta- \Delta_2} |x_{23}|^{ \Delta_2 + \theta - \Delta_1 }}\,,
\end{equation}
where $\theta=\Delta_j-j$ is the \emph{twist}\index{twist} (dimension minus spin) of the operator $\hat{\op}_{j}$,
\begin{equation}
\hat{Y}_{12,3}=Y^\mu z_\mu \,, \qquad Y^\mu (x_{13},x_{23}) = \frac{x_{13}^\mu}{x_{13}^2} - \frac{x_{23}^\mu}{x_{23}^2} = \frac{1}{2}\partial_{x_3}^\mu \ln \left(\frac{x_{23}^2}{x_{13}^2}  \right)
\end{equation}
and where $x_{ij}^\mu=x_i^\mu-x_j^\mu$. Note, that the dimension of the twist operator and the structure constants both acquire corrections  in perturbation theory
\begin{align}
\theta = \Delta_j - j &= 2 + \gamma_j(g^2),\qquad
C_{\op \tilde{\op} j}(g^2)  = C_{\op \tilde{\op} j}^{(0)} + g^2 C_{\op \tilde{\op} j}^{(1)} + \op(g^4)\,.
\end{align}

In order to carry out the calculations we evaluate the correlator in a soft-limit, where the momentum coming in at the spin $j$ operator becomes zero. In position space this corresponds to an integration over the corresponding point $x_3$
\begin{equation}\label{eqn:limit-procedure}
\int \frac{d^dp}{(2\pi)^d} e^{i p \cdot x_{12}} \langle \op(p) \tilde{\op}(-p) \hat{\mathbb{O}}_j(0) \rangle = \int d^dx_3  \langle \op(x_1) \tilde{\op}(x_2) \hat{\mathbb{O}}_j(x_3) \rangle\,.
\end{equation}

This limit largely simplifies the perturbative calculation, since all three-point diagrams effectively reduce to two-point diagrams and the dependence on the one-loop mixing matrix drops out completely, as we will see in section \ref{sec:calculation}. We find that the correction to the structure constant takes the simple form
\begin{equation}\label{eqn:normalised-structure-constants}
C^\prime_{\op  \tilde{\op} j}(g) = {C^\prime}_{\op \tilde{\op} j}^{(0)} \left(1 +   \frac{g^2N}{8 \pi^2}  \left(2H_j(H_j -  H_{2j})- H_{j,2}\right) + \op(g^2) \right)\,,
\end{equation} 
where $H_{j,m}$ are  generalised harmonic sums $H_j=\sum_n^j 1/n$, $H_{j,r}=\sum_n^j 1/n^r$.
The result is in agreement with the extraction of the structure constants from the analysis of the operator product expansion (OPE) of four-point functions of half-BPS operators \cite{Dolan:2004iy}. Recently, the OPE and the three-loop expression  for the four-point correlator of half-BPS operators \cite{Eden:2011we} were used to extract the structure constants up to three loops \cite{Eden:2012rr}. Our calculation thus directly confirms the results obtained from the OPE.

In the following sections we will only summarise the main steps of our calculation in order to keep the presentation short, more details will be made available in  \cite{wiegandt:2012phdthesis}.

\section{Three-Point Functions of Twist Operators}\label{sec:calculation}
We use the mostly minus metric and the conventions as given in Appendix \ref{sec:conventions} and adapt the notation from  \cite{Belitsky:2007jp}.
The renormalised twist-two operators are given by
\begin{equation}\label{equ:renormalised-op}
 \hat{\mathbb{O}}_j(x) = \sum_k \mathbb{Z}_{jk} \hat{\partial}^{j-k} \hat{\op}_k(x)\,,
\end{equation}
where the tree-level correlator is
\begin{equation}
\hat{\op}_j (x) = \sum_{k=0}^j a_{jk}^{1/2}\,\tr\left( \hat{D}^k \phi^{12}(x) \hat{D}^{j-k} \phi^{12}(x)\right)\,,
\end{equation}
the coefficients $a_{jk}^{1/2}$ are related to the Gegenbauer polynomials as in \eqref{eqn:relation-gegenbauer-polynomials} and the one-loop mixing matrix in the \emph{conformal scheme} has the form
\begin{equation}\label{eqn:mixing-matrix}
\mathbb{Z}_{jk}= \delta_{jk} + g^2\mathbb{Z}_{jk}^{(1)}+ \op(g^4) =\delta_{jk} + g^2\left(-B_{jk}^{(1)}+\frac{1}{\epsilon}\delta_{jk} Z_{j}^{(1)}\right) + \op(g^4)\,.
\end{equation}
At one-loop level the divergent part of the renormalisation matrix is diagonal with $Z_j^{(1)} = H_j/(4\pi^{d/2}) \exp( \epsilon \gamma_E)$ and determines the anomalous dimension via $\gamma_j = -\mu d/d\mu \ln Z_j (\mu)$.  
The finite matrix $B_{jk}(g^2)$  diagonalizes the anomalous dimension matrix at higher-loop level $\gamma_j (g^2)\delta_{jk}= (B^{-1} \gamma B)_{jk}$ and
accounts for non-diagonal contributions to the two-point functions at loop-level. We do not specify $B_{jk}$, since it drops out in the limit that we consider.
The operators given by \eqref{equ:renormalised-op} then have diagonal, conformal two-point functions also at loop-level and are fixed up to the normalization $C_j(g^2)$
\begin{equation}
 \langle \hat{\mathbb{O}}_j(x_1) \hat{\mathbb{O}}_k(x_2) \rangle = \delta_{jk} C_j(g^2) \frac{( \hat{I}_{12})^j}{(-x_{12}^2)^{2j+\theta}}\qquad (j~ \text{even})\,,
\end{equation}
where $\theta=\Delta_j - j = 2 +\gamma_j(g^2)$ and $I^{\mu\nu}_{12}= \eta^{\mu\nu} -2 x_{12}^\mu x_{12}^\nu / x_{12}^2$. In order to read-off the structure constants in the limit \eqref{eqn:limit-procedure} we have to integrate \eqref{eqn:three-point-structure} over $x_3$, which yields
\begin{equation}\label{eqn:renormalized-three-point-function}
 \int d^dx_3 \langle \op \tilde{\op} \hat{\mathbb{O}}_j\rangle = N(g^2) \left(C_{\op\tilde{\op}j}^{(0)}+g^2 C_{\op\tilde{\op}j}^{(1)} \right)\frac{(\hat{x}_{12})^j}{(-x_{12}^2)^{j+d-3+\gamma_j(g^2)/2}} 
\end{equation}
where $N(g^2)$ is a normalisation factor explicitly given in \eqref{eqn:N(g^2)-for-d-dimensions}. By calculating the left-hand side in the limit $p_1+p_2\to0$ one can thus easily read-off the structure constants, after Fourier transforming the momentum space expression\footnote{
An analogue of this procedure appears in \cite{Costa:2010rz}, where a deformation of the N = 4 SYM Lagrangian
by integrated local operators is considered and two-point functions of the deformed theory are studied at leading
order in the deformation parameter.}

\section{Tree-Level Calculation}

In momentum space, the tree-level three-point function with the twist-operator in the representation \eqref{eqn:definition-with-gegenbauer-polynomials} reads
\begin{align}\label{eqn:tree-expression}
\langle \op(p_1) \tilde{\op}(p_2) \hat{\op}_j \rangle &= \frac{1}{4} \, i^{3+j}g^6 \delta^{aa} \int \frac{d^dk}{(2\pi)^d} \frac{(\hat{p}_1+\hat{p}_2)^j C_j^{1/2}\left( \frac{2\hat{k}-\hat{p}_1-\hat{p}_2}{\hat{p}_1+\hat{p}_2} \right)}{k^2 (p_1-k)^2(p_1+p_2-k)^2}\,,
\end{align}
Now we take the limit $p_1+p_2 \to 0$ in momentum space. 
Then, due to the factor $(p_1+p_2)^j$ only the term with the highest power, i.e. $j$, in the Gegenbauer polynomial can survive. The corresponding coefficient reads
\begin{equation}\label{eqn:highest-power-gegenbauer}
c_{jj}^{1/2}  = 2^{1-j} \frac{\Gamma(2j)}{\Gamma(j)\Gamma(j+1)} \quad \text{where} \quad C_j^{1/2} (x) = \sum_{k=0}^j c^{1/2}_{jk} x^k \,.
\end{equation}
Thus, in this limit the three-point integral in \eqref{eqn:tree-expression} becomes a two-point integral with a \emph{doubled} propagator 
\begin{equation}
 \int \frac{d^dk}{(2\pi)^d} \frac{(\hat{p}_1+\hat{p}_2)^j C_j^{1/2}\left( \frac{2\hat{k}-\hat{p}_1-\hat{p}_2}{\hat{p}_1+\hat{p}_2} \right)}{k^2 (p_1-k)^2(p_1+p_2-k)^2} \to  2^j c_{jj}^{1/2}\int \frac{d^dk}{(2\pi)^d} \frac{(k)^j }{k^4 (p_1-k)^2}\,,
\end{equation}
which can easily be solved using \eqref{eqn:bubble-integral} in the appendix.
Fourier transformation to $x$-space using \eqref{eqn:result-fourier-trafo-mink} and insertion of the coefficients yields
\begin{align}\label{eqn:tree-level-three-point-function-in-x-space-calculated}
\int d^dx_3 \langle \op \tilde\op \hat{\op}_j\rangle  
&= -ig^6 \delta^{aa} \frac{2^{d-7+j}}{\pi^{\frac{d}{2}}}\frac{\Gamma(2j)\Gamma(h-1)\Gamma(j-2+h)}{\Gamma(j)\Gamma(j+1)} \frac{(\hat{x}_{12})^j}{(-x_{12}^2)^{2h-3+j}}\,.
\end{align}
Comparing with \eqref{eqn:renormalized-three-point-function} we can directly read off the tree-level structure constant for $d=4$ and arbitrary even spin $j$
\begin{align}\label{eqn:tree-level-strcuture-constant-N=4}
C_{\op\tilde{\op}j}^{(0)} = g^6 \delta^{aa} \frac{2^{j-8}}{\pi ^6} \Gamma (j+1)\,.
\end{align}

\section{One-Loop Calculation}
\begin{figure}[h]
\centering
\subfloat[]{\begin{minipage}[c]{2.3cm}
		~\includegraphics[width=1 \textwidth]{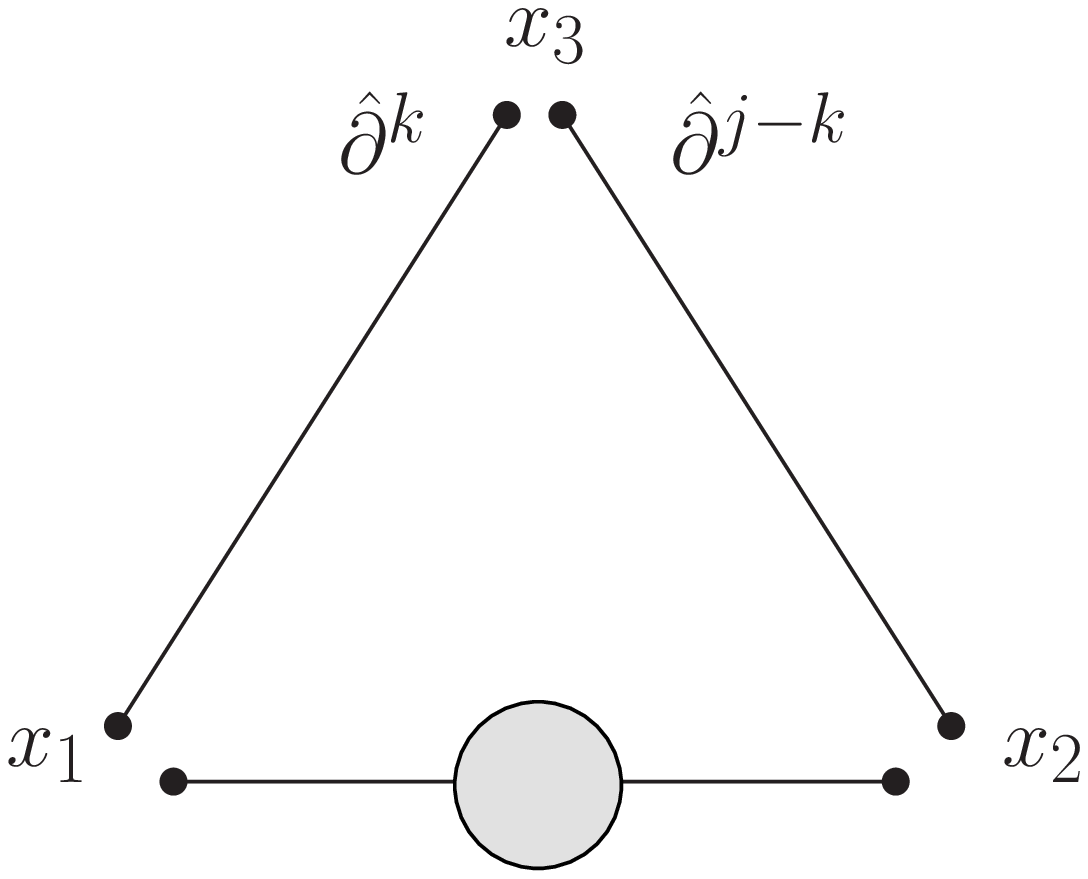}\label{fig:3point1loopb}
		\end{minipage}
		
}~~
\subfloat[]{\begin{minipage}[c]{2cm}
		~\includegraphics[width=1 \textwidth]{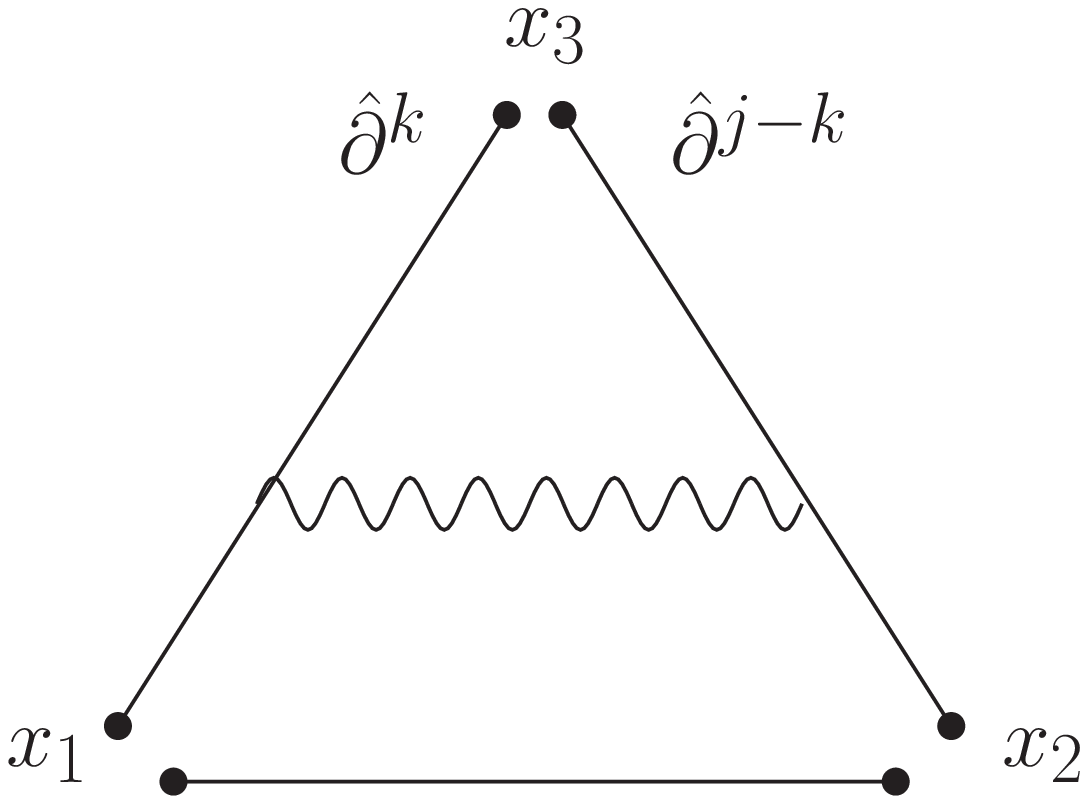}\label{fig:3point1loopg}
		\end{minipage}
		
}~~
\subfloat[]{\begin{minipage}[c]{2cm}
		~\includegraphics[width=1 \textwidth]{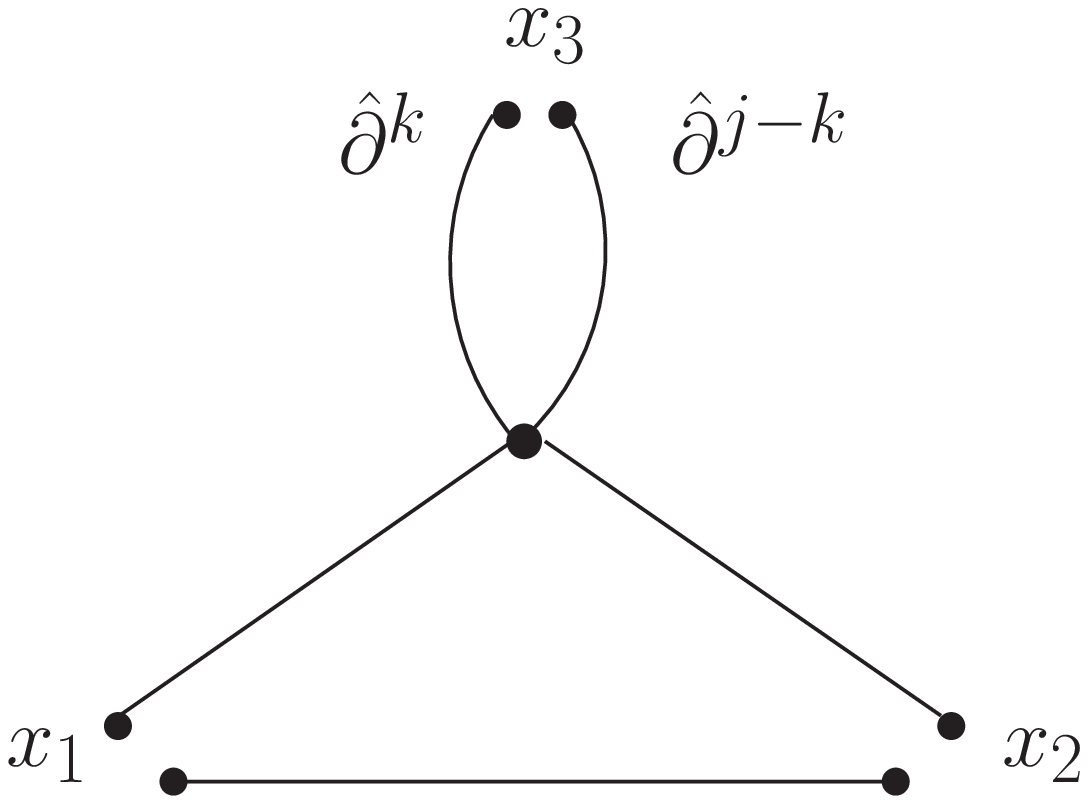}\label{fig:3point1loopl}
		\end{minipage}
		
}~~
\subfloat[]{\begin{minipage}[c]{2cm}
		~\includegraphics[width=1 \textwidth]{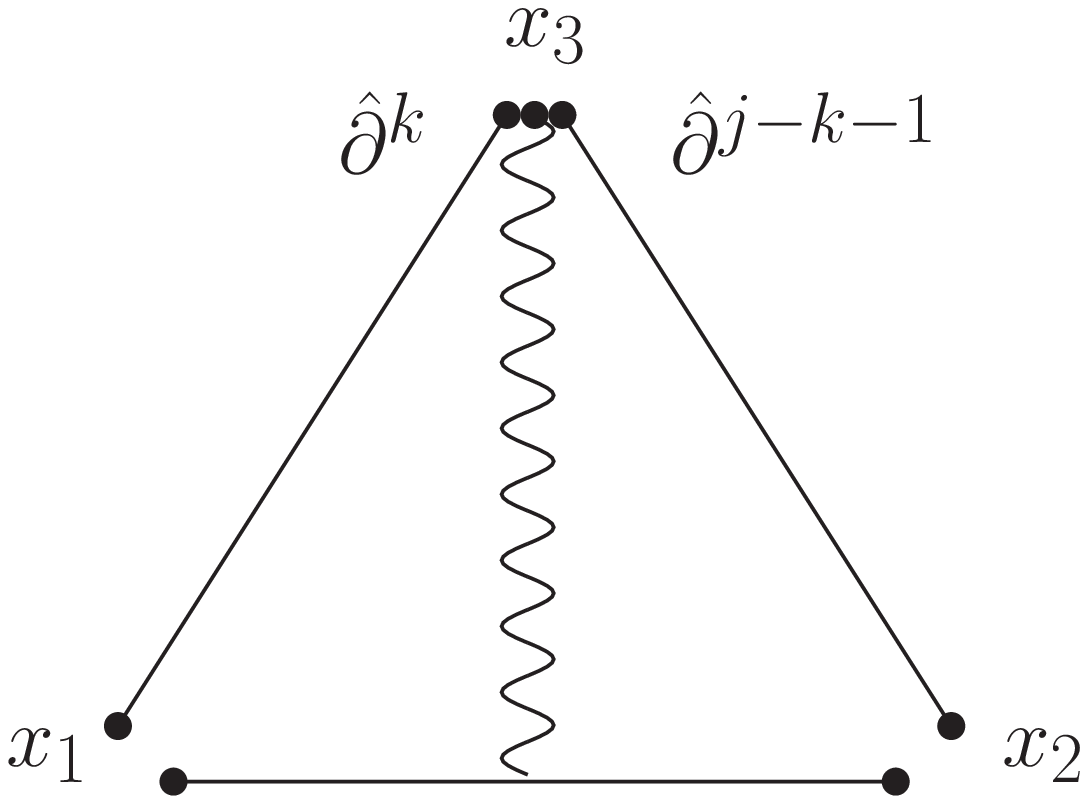}\label{fig:3point1loopf}
		\end{minipage}
		
}~~
\\
\subfloat[]{\begin{minipage}[c]{2cm}
		~\includegraphics[width=1 \textwidth]{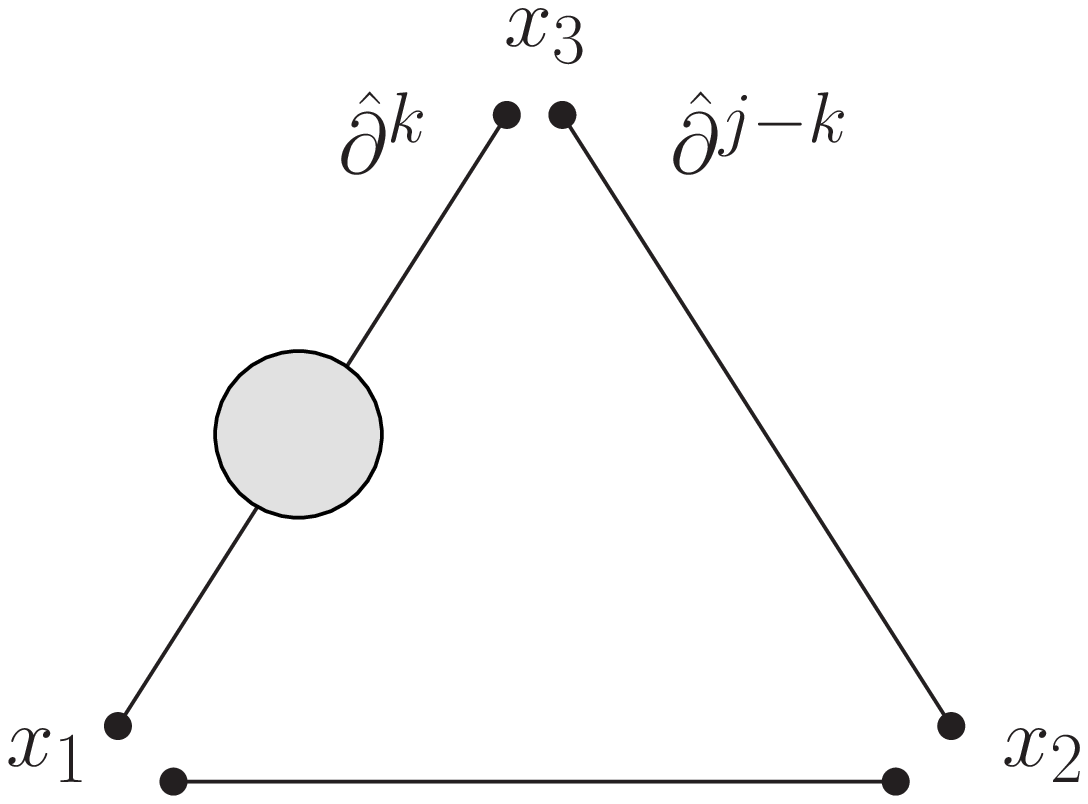}\label{fig:3point1loopa}
		\end{minipage}
		
}~~
\subfloat[]{\begin{minipage}[c]{2cm}
		~\includegraphics[width=1 \textwidth]{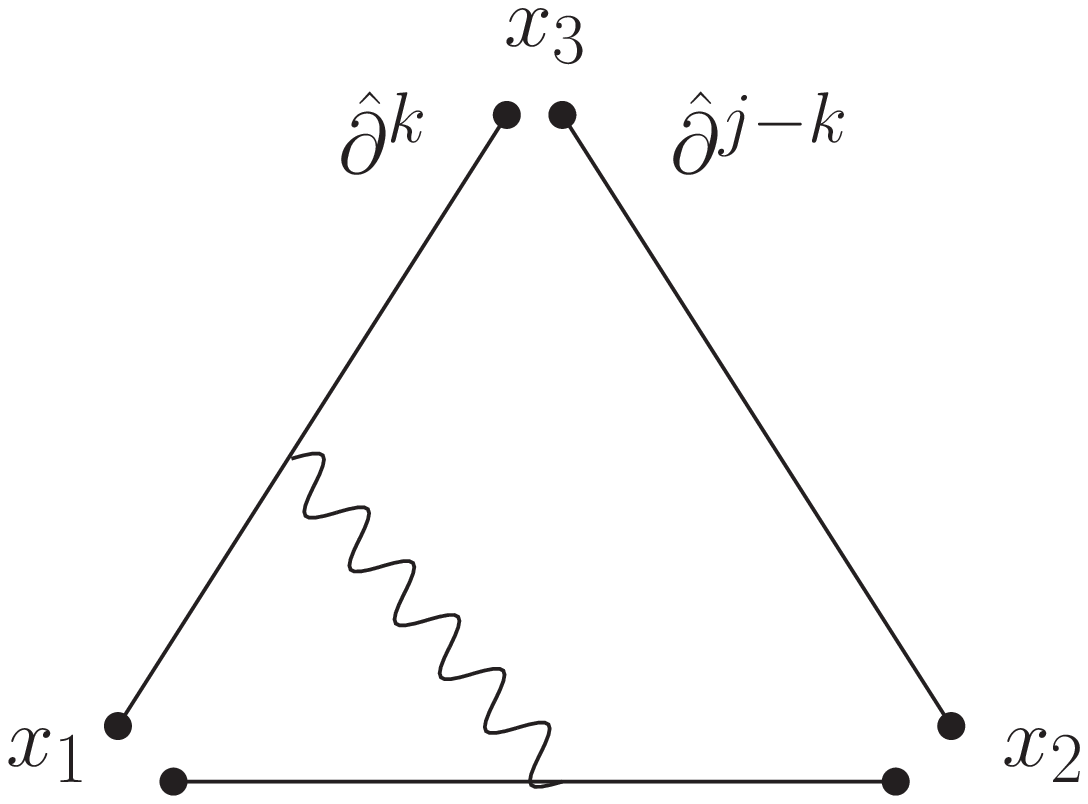}\label{fig:3point1looph}
		\end{minipage}
		
}~~
\subfloat[]{\begin{minipage}[c]{2cm}
		~\includegraphics[width=1 \textwidth]{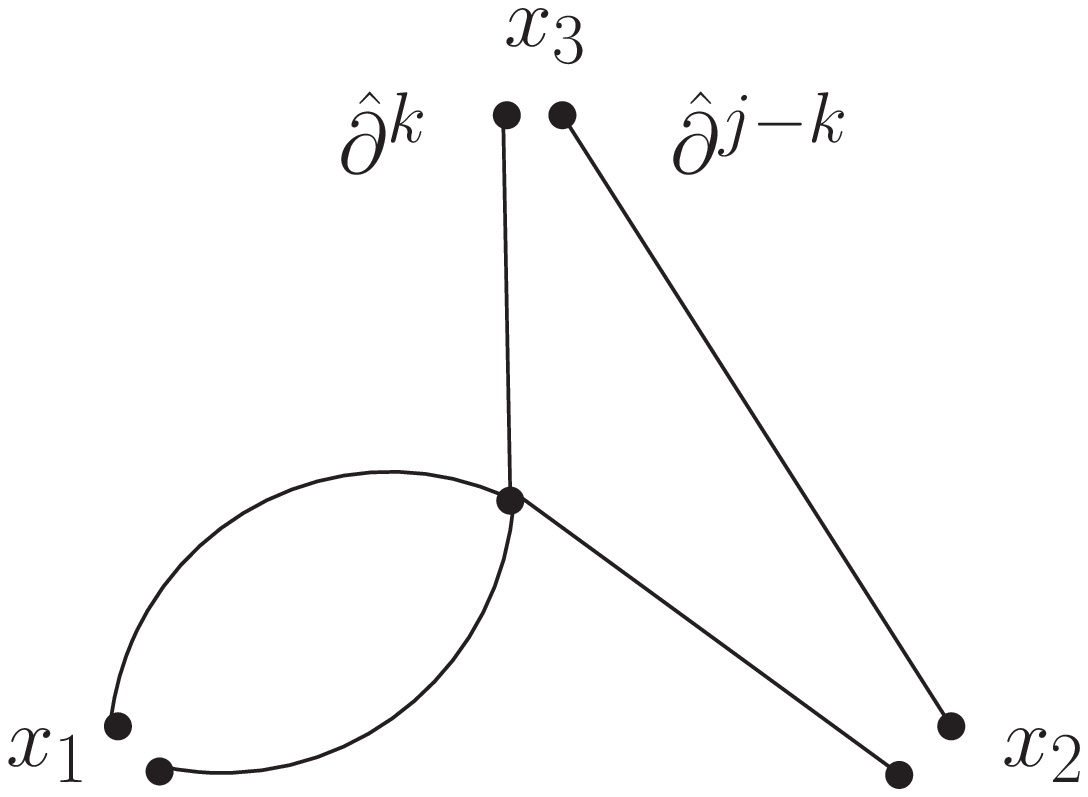}\label{fig:3point1loopj}
		\end{minipage}
		
}~~
\subfloat[]{\begin{minipage}[c]{2cm}
		~\includegraphics[width=1 \textwidth]{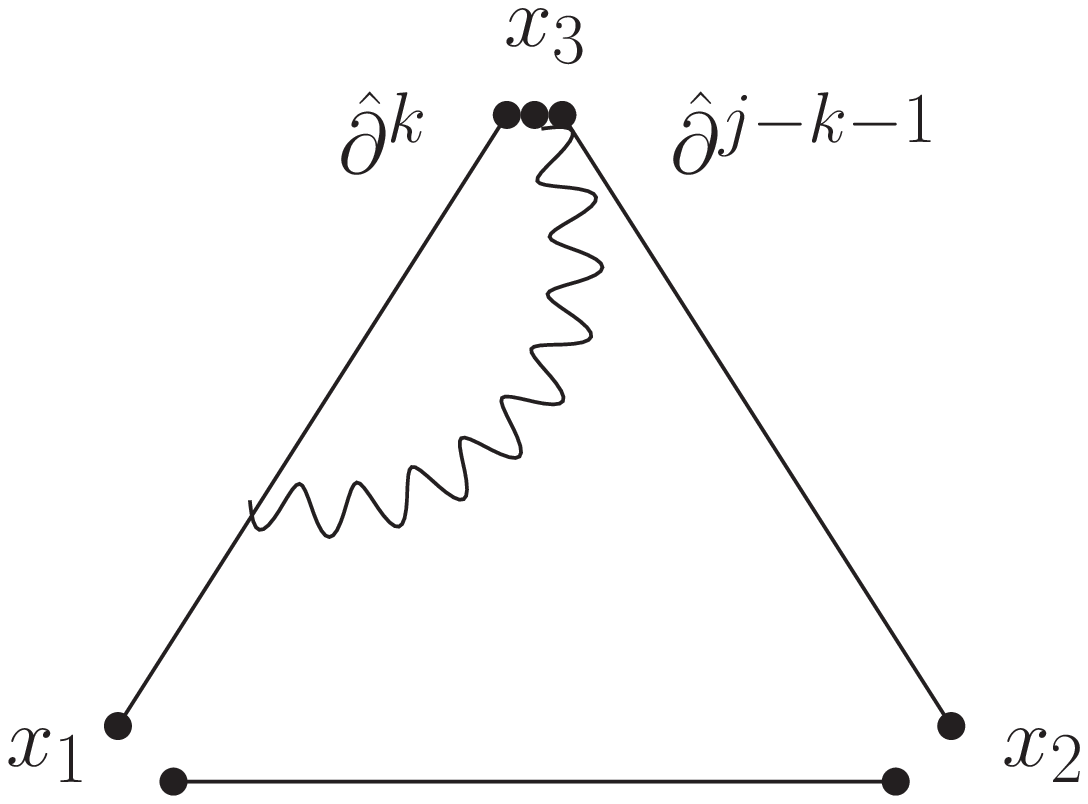}\label{fig:3point1loope}
		\end{minipage}
}~~
\\
\subfloat[]{\begin{minipage}[c]{2cm}
		~\includegraphics[width=1 \textwidth]{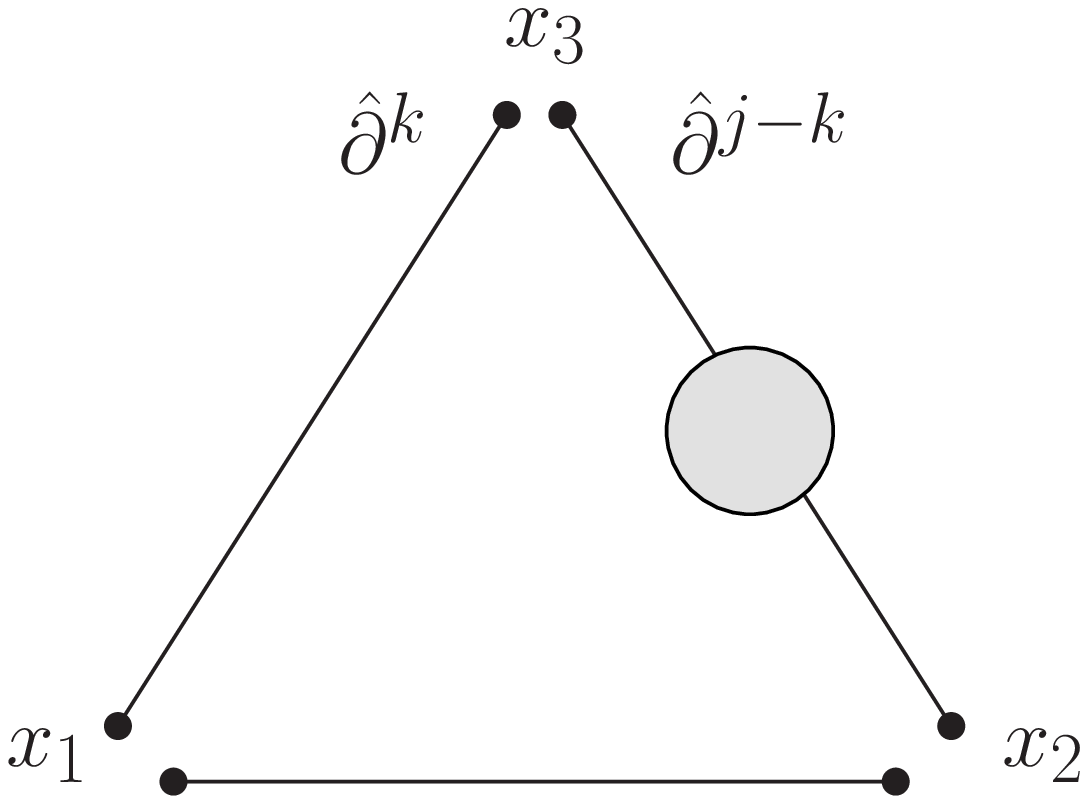}\label{fig:3point1loopc}
		\end{minipage}
		
}~~
\subfloat[]{\begin{minipage}[c]{2cm}
		~\includegraphics[width=1 \textwidth]{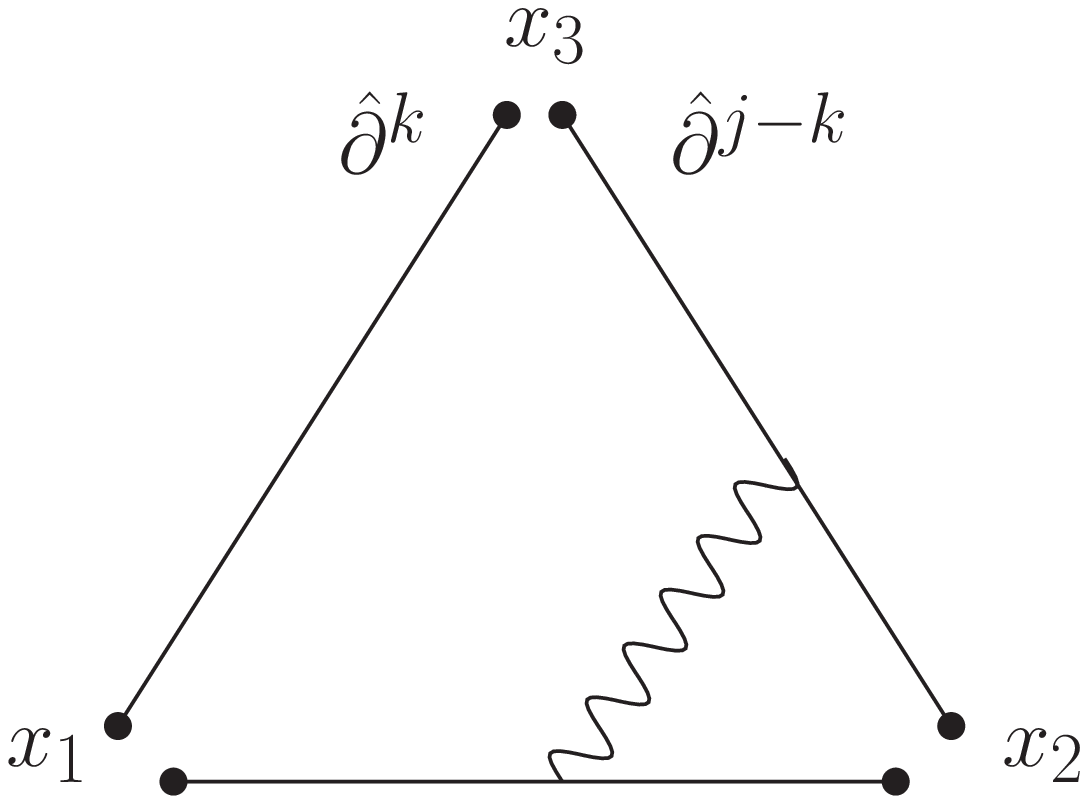}\label{fig:3point1loopi}
		\end{minipage}
		
}~~
\subfloat[]{\begin{minipage}[c]{2cm}
		~\includegraphics[width=1 \textwidth]{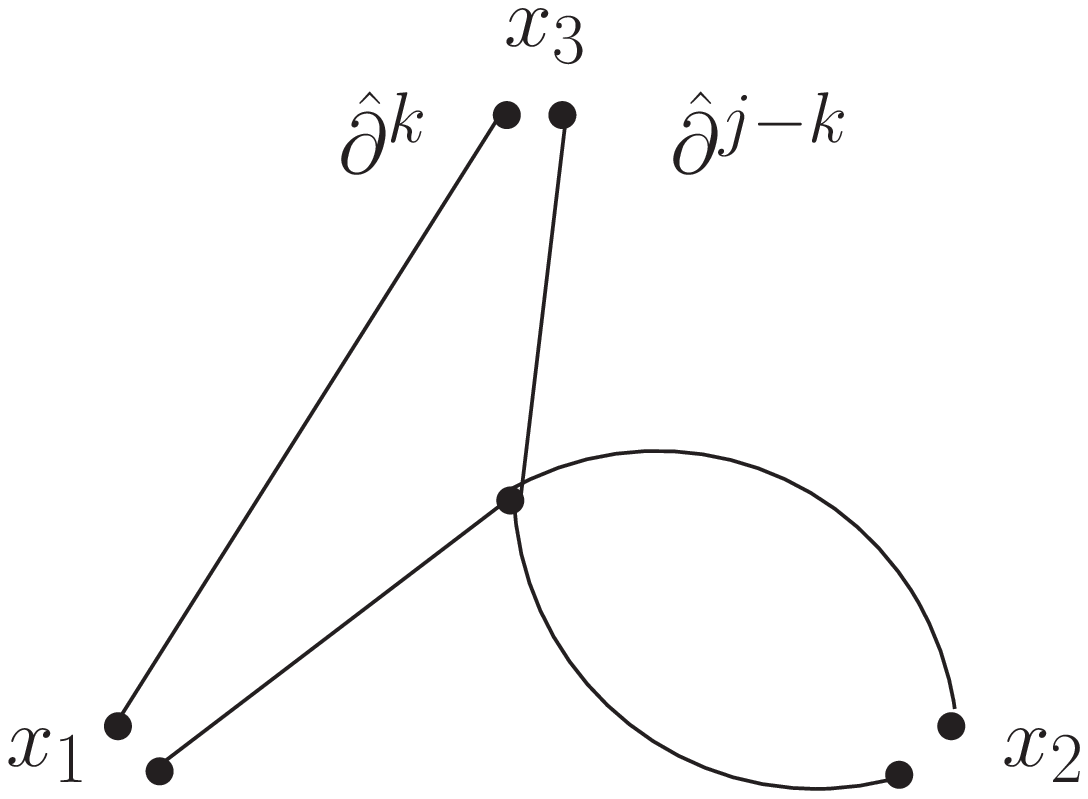}\label{fig:3point1loopk}
		\end{minipage}
}~~
\subfloat[]{\begin{minipage}[c]{2cm}
		~\includegraphics[width=1 \textwidth]{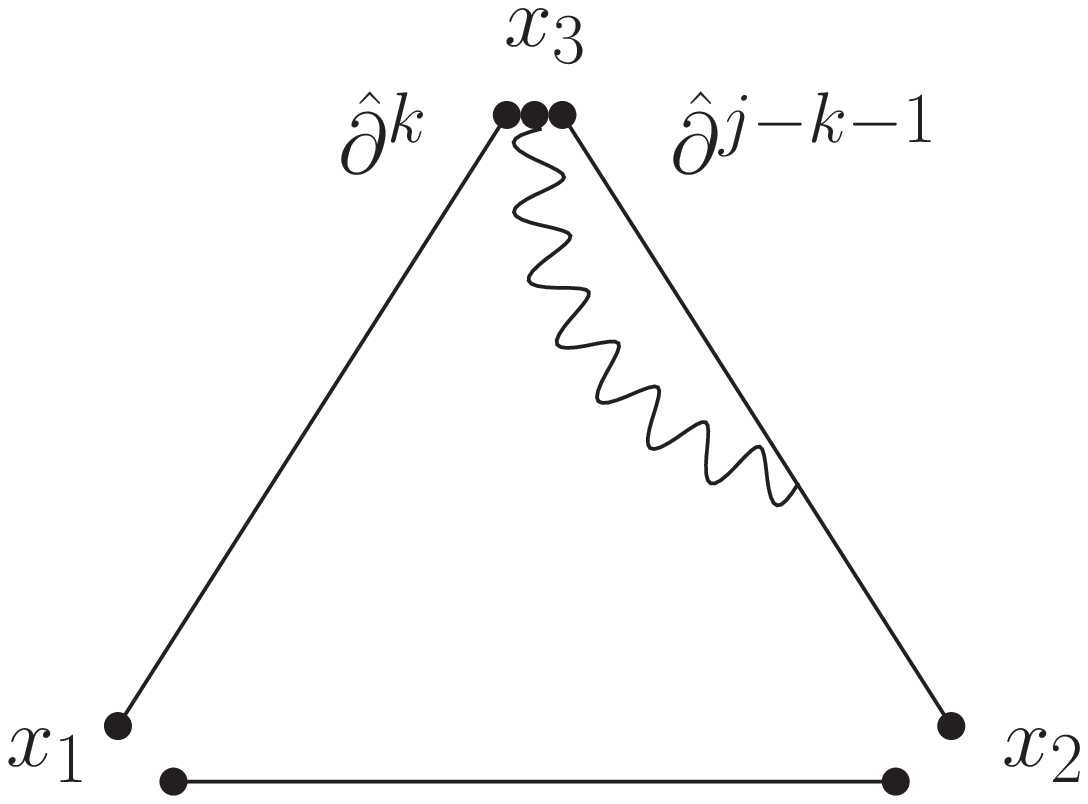}\label{fig:3point1loopd}
		\end{minipage}
		
}~~
\caption{Feynman diagrams contributing to the three-point function at one loop. For $j$ even the diagrams in the second and third row are identical.}
\label{fig:three-point function at 1-loop}
\end{figure}

In order to read-off the one-loop structure constant we have to compute the renormalised three-point function using the renormalised operator given by  \eqref{equ:renormalised-op}-\eqref{eqn:mixing-matrix}. From
\begin{equation}\nn
\int d^dx_3 \langle \op \tilde\op \hat{\mathbb{O}}_j \rangle= \sum_k \mathbb{Z}_{jk} \int d^dx_3 \hat{\partial}_3^{j-k} \langle \op \tilde\op \hat{\op}_k \rangle= \sum_k \mathbb{Z}_{jk} \delta_{jk} \int d^dx_3\langle \op \tilde\op \hat{\op}_k \rangle
\end{equation}
it is however immediately clear, that the finite matrix $B_{jk}$ drops out in the limit that we consider, which vastly simplifies the calculation. As can be seen from \eqref{eqn:mixing-matrix}, we then only need to calculate the one-loop diagrams and subtract the tree-level contribution\footnote{It needs to be evaluated including $\op(\epsilon)$ terms, because it is multiplied with $1/\epsilon$.} multiplied with the renormalisation constant $Z_j$. The one-loop diagrams that enter the calculation are shown in figure \ref{fig:three-point function at 1-loop}.

It is instructive to write down all diagrams before employing the limit $p_1+p_2 \to 0$ and to convince oneself that diagrams which could lead to ambiguities, such as the diagram in figure \ref{fig:3point1loopj}, which vanishes in dimensional regularisation when taking the limit, cancel with contributions from other diagrams. 

\begin{figure}[h]
\center
 \includegraphics[width=.08\textwidth]{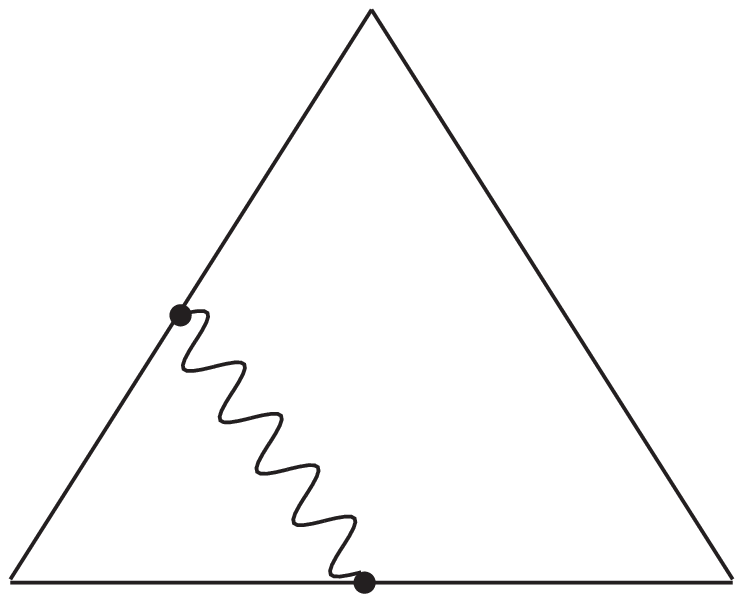} 
\raisebox{.35cm}{$= -2p_1^2$}
 \includegraphics[width=.08\textwidth]{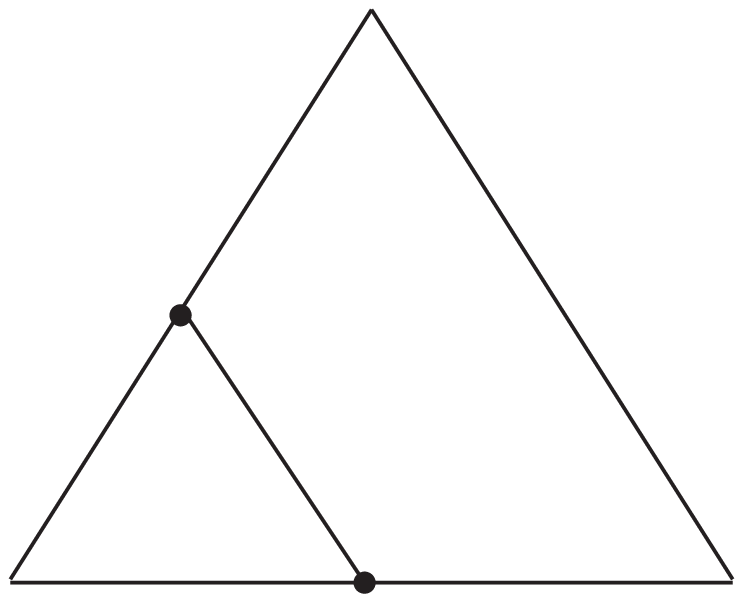} 
   \raisebox{.35cm}{$ -$}
 \includegraphics[width=.08\textwidth]{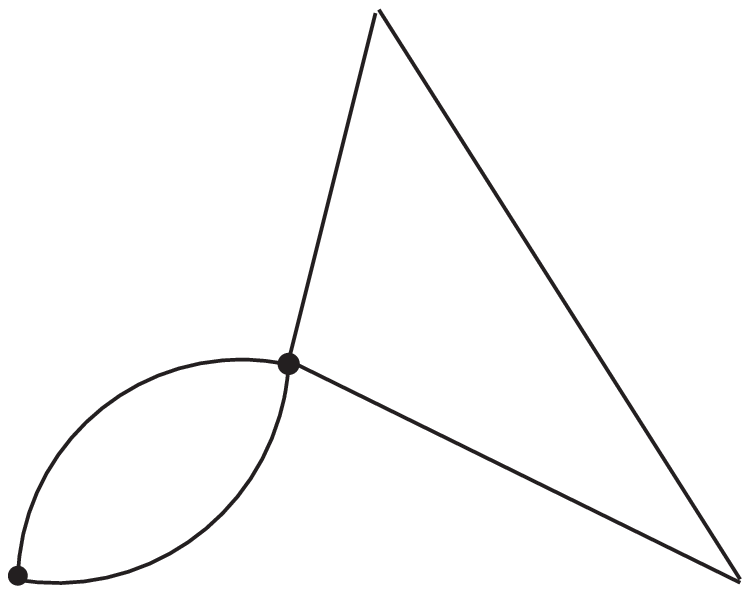} 
   \raisebox{.35cm}{$ +$}
 \includegraphics[width=.08\textwidth]{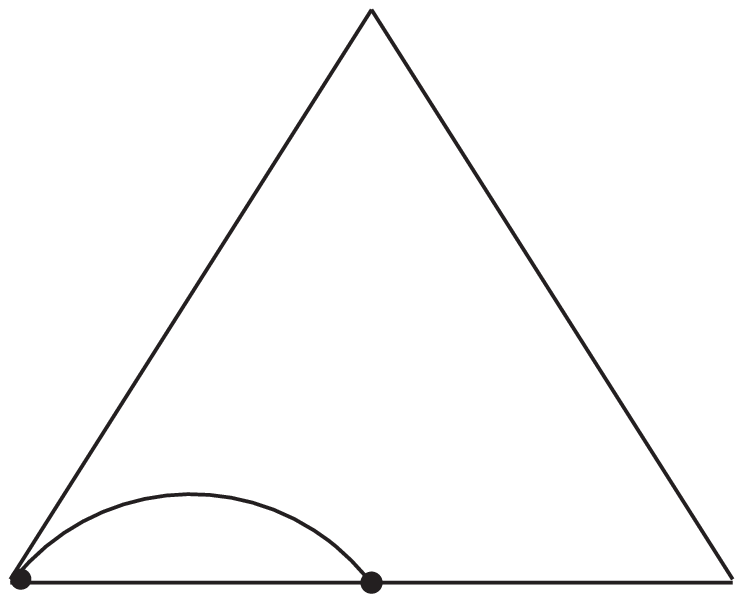} 
  \raisebox{.35cm}{$ +$}
 \includegraphics[width=.08\textwidth]{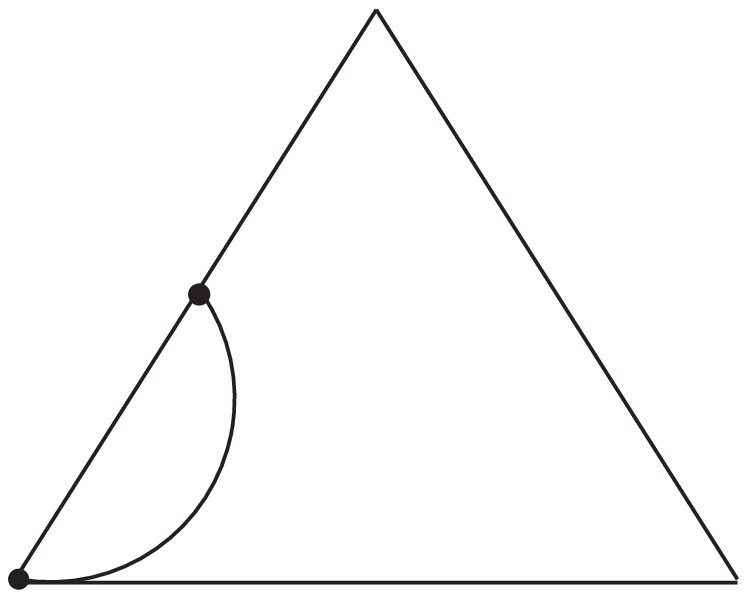} 
 \raisebox{.35cm}{$ +$}
 \includegraphics[width=.08\textwidth]{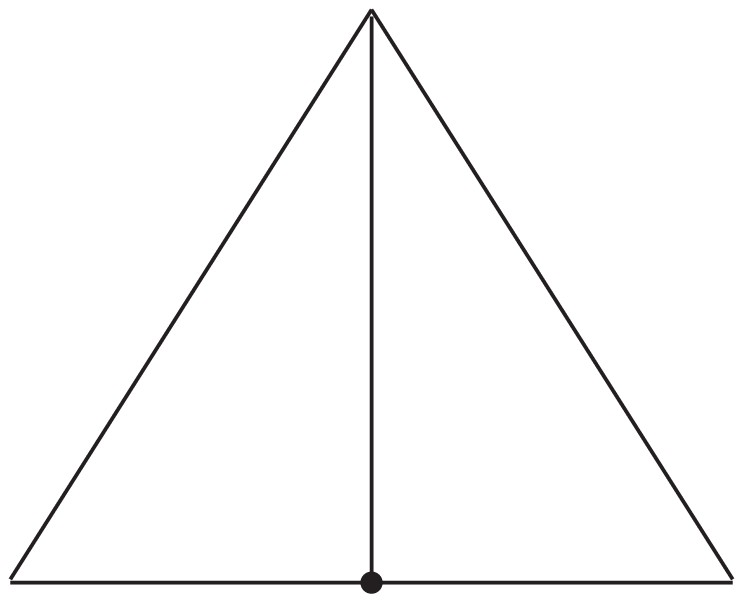} 
  \raisebox{.35cm}{$ +$}
 \includegraphics[width=.08\textwidth]{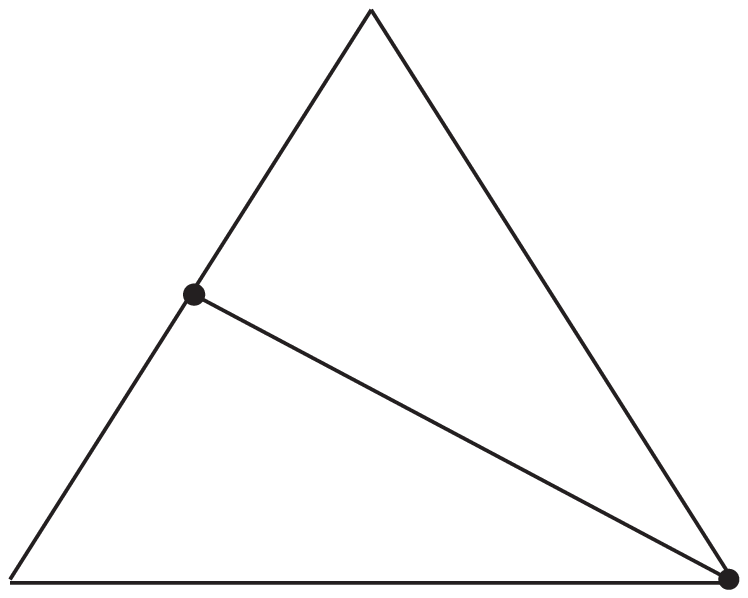} 
\caption{Due to the momenta from the gluon vertices, the integrand decomposes into simpler integrals, which cancel with the self-energy and four-scalar interaction terms.}
\label{fig:decomposition-gluon-diagrams}
\end{figure}

Indeed, all diagrams with a scalar four-point interaction vertex, i.e. fig. \ref{fig:3point1loopl}, \ref{fig:3point1loopj}, \ref{fig:3point1loopk} cancel against contributions from fig. \ref{fig:3point1loopg}, \ref{fig:3point1looph}, \ref{fig:3point1loopi}, which becomes clear when considering the decomposition of diagrams with two gluon vertices into scalar integrals, see figure \ref{fig:decomposition-gluon-diagrams}.

 For $j=0$, the diagrams in the last column do not appear, since there is no covariant derivative that the gauge field at $x_3$ could arise from and half of the diagrams \ref{fig:3point1loopa}, \ref{fig:3point1loopb} cancels against a contribution from \ref{fig:3point1looph}, half of the diagrams \ref{fig:3point1loopb}, \ref{fig:3point1loopc} cancels against a contribution from \ref{fig:3point1loopi} and half of the diagrams \ref{fig:3point1loopa}, \ref{fig:3point1loopc} cancels against a contribution from \ref{fig:3point1loopg}. The remaining finite contributions from \ref{fig:3point1loopg}, \ref{fig:3point1looph}, \ref{fig:3point1loopi} cancel due to a relation between scalar three-point integrals given in \cite{Usyukina:1994iw}. Thus for $j=0$ all contributions exactly cancel, which is due to the fact that in this case all of the operators are protected.

For $j \neq 0$ the above cancellations between divergent diagrams where the divergence  is located at the points $x_1$ resp. $x_2$, i.e. where the BPS operators sit, remain true as one would expect\footnote{Would divergences at these points remain, the operators acquired an anomalous dimension, we do however have protected operators with vanishing anomalous dimension at $x_1,x_2$.}. The cancellations between half of the diagrams \ref{fig:3point1loopa}, \ref{fig:3point1loopc} with  a contribution from \ref{fig:3point1loopg} does however not take place anymore, and the spin-$j$ operator at $x_3$ acquires an anomalous dimension $\gamma_j$. 
For the remaining diagrams in the limit $p_1+p_2$, for the same reason as in the tree-level calculation, we only need the coefficient $c_{jj}^{1/2}$ with the highest power of the Gegenbauer polynomial that was given in \eqref{eqn:highest-power-gegenbauer}. Since $p_1=-p_2=p$ and $j$ is even we find the same result for all diagrams which are equal under $p_1 \leftrightarrow p_2$. 
The integrals turn into simple two-point integrals that can be easily solved using \eqref{eqn:bubble-integral}. Denoting from now on $p_1=p$ we find
\begin{align}\label{eqn:diagram-g-plus-j}
\langle \op\tilde{\op} & \hat{\op}_j \rangle^{\ref{fig:3point1loopg}} +\langle  \op\tilde{\op} \hat{\op}_j \rangle^{\ref{fig:3point1loopl}} = \\ \nn 
&= - c_{jj}^{1/2} 2^{j-1} i^{2+j}  g^8 N \delta^{aa}  b_j(4-\frac{d}{2},1)\left(b_j(2,1)+b_j(1,1) \right) \frac{\hat{p}^j}{(-p^2)^{5-d}}\,.
\end{align}
where $b_n(\alpha_1,\alpha_2)$ is defined in \eqref{eqn:bna1a2}.
As mentioned before, half of the self-energy diagrams in fig. \ref{fig:3point1loopa}, \ref{fig:3point1loopc} do not cancel for $j \neq 0$ and the expression is
\begin{align}\label{eqn:diagram-a-in-limit}
\frac{1}{2} \langle \op\tilde{\op} \hat{\op}_j \rangle^{\ref{fig:3point1loopa}} & + \frac{1}{2} \langle \op\tilde{\op} \hat{\op}_j \rangle^{\ref{fig:3point1loopc}} = \\ \nn
&\quad \quad \quad= c_{jj}^{1/2} i^{2+j}  g^8 N \delta^{aa}  2^{j-1} b_0(1,1) b_j(4-\frac{d}{2},1) \frac{\hat{p}^j}{(-p^2)^{5-d}}\,.
\end{align}
Applying the limit to the diagrams where the divergence is located at $x_1$ we find
\begin{align}\label{eqn:diagram-l-j-b-c}
\langle \op \tilde{\op} \hat{\op}_j \rangle^{\ref{fig:3point1looph}} &+ \langle \op \tilde{\op} \hat{\op}_j \rangle^{\ref{fig:3point1loopj}} + \frac{1}{2}\langle \op \tilde{\op} \hat{\op}_j \rangle^{\ref{fig:3point1loopa}} + \frac{1}{2}\langle \op \tilde{\op} \hat{\op}_j \rangle^{\ref{fig:3point1loopb}} \\ \nn 
& \quad =  i^{2+j}  g^8 N \delta^{aa}  2^{j-2} c_{jj}^{1/2} \frac{\hat{p}^j}{(-p^2)^{d-5}} \\ \nn
&\qquad \left( 2c_{0j}(1,1,1,2,1) + c_{0j}(1,1,1,1,1)+ c_{0j}(1,1,1,2,0) \right) \,.
\end{align}
The integrals $c_{nm}(a_1,a_2,a_3,a_4,a_5)$ are defined in \eqref{eqn:definition-cmn} and solved\footnote{These integrals were considered before in \cite{Kazakov:1986mu}.} in Appendix \ref{sec:integrals}  using the IBP technique.
All these integrals are finite. The diagrams where the divergence is located at $x_2$, i.e. \ref{fig:3point1loopk}, \ref{fig:3point1loopi}, \ref{fig:3point1loopb}, \ref{fig:3point1loopc} yield the same result, since they are related to the ones above by $p_1 \leftrightarrow p_2$.

For $j\neq 0$ we have additional diagrams with exactly one gauge field from the covariant derivative in $\hat\op_j$. 
The diagrams are shown in figure \ref{fig:3point1loopf}, \ref{fig:3point1loope}, \ref{fig:3point1loopd}. 
Diagrams \ref{fig:3point1loope} and \ref{fig:3point1loopd} are  equal in our limit.
The appearing integrals are simple two-point integrals and we find
\begin{align}\label{eqn:diagram-7-d-in-limit}\nn
\langle \op(p) \tilde{\op}(-p) & \hat{\op}_j(0) \rangle^{\ref{fig:3point1loope}} = - i^{6+j} \frac{ g^8 N \delta^{aa} }{4} b_j(4-d/2,1)\sum_{k=1}^j a_{jk}^{1/2} \sum_{m=1}^{k}  \sum_{n=0}^{m-1}{k \choose m} {m-1 \choose n}(-1)^m \\ 
&  \frac{\hat{p}^j}{(-p^2)^{5-d}}   \left[ b_{k-n}(1,1)-b_{k-n-1}(1,1) +b_{j-k+n}(1,1)-b_{j-k+n+1}(1,1)  \right]\,.
\end{align}
and one can proceed to solve the sums.
Diagram  \ref{fig:3point1loopf} is calculated in a very similar way and reads
\begin{align}\label{eqn:diagram-7-f-in-limit} \nn
\langle \op(p) \tilde{\op}(-p) & \hat{\op}_j(0) \rangle^{\ref{fig:3point1loopf}}  = i^{6+j} \frac{ g^8 N \delta^{aa} }{2} \sum_{k=1}^j a_{jk}^{1/2} \sum_{m=1}^k \sum_{n=0}^{m-1} {k \choose m} {m-1 \choose n}(-1)^{k-m+n} \\
& \frac{\hat{p}^j}{(-p^2)^{5-d}} \Big( 2 c_{k-1-n,j-k+n}+c_{k-n,j-k+n}+c_{k-1-n,j-k+n+1}\Big)\,,
\end{align}
where all integrals $c_{nm}=c_{nm}(1,1,1,1,1)$ are finite and the solution can be found in appendix \ref{sec:integrals}.

\subsection{Full Bare Three-Point Function}
Taking into account the exact cancellations between diagrams, as well as the fact that diagrams with $p_1 \leftrightarrow p_2$ are identical in the limit $p_1\to -p_2$ for $j$ even, the  bare one-loop contribution to the three-point function is given by
\begin{align}\label{eqn:bare-three-point-function}
\langle \op \tilde{\op} \hat{\op}_j \rangle^{(1)} &= \sum_{\alpha= a..l} \langle \op \tilde{\op} \hat{\op}_j \rangle^{\ref{fig:three-point function at 1-loop}\alpha} \\ \nn
&= \eqref{eqn:diagram-g-plus-j}+  \eqref{eqn:diagram-a-in-limit} + 2  \times  \eqref{eqn:diagram-l-j-b-c} +2   \times \eqref{eqn:diagram-7-d-in-limit} +   \eqref{eqn:diagram-7-f-in-limit} \,.
\end{align}
We have calculated all diagrams in momentum space. In order to read off the structure constant we Fourier transform the expression to position space using \eqref{eqn:result-fourier-trafo-mink}.

As a check of the calculation one can extract the divergent part of \eqref{eqn:bare-three-point-function} and read off the anomalous dimension from the three-point function. As mentioned before, only divergences located at $x_3$ remain and we get the following contributions to the anomalous dimension
\begin{equation}\nn
\gamma_j^{(\ref{fig:3point1loopa}+\ref{fig:3point1loopc})/2}  = \frac{g^2N}{4\pi^2},~~   \gamma_j^{\ref{fig:3point1loopg}+\ref{fig:3point1loopl}}  = \frac{g^2N}{4\pi^2}\left(-\frac{1}{j+1}\right),~~  \gamma_j^{\ref{fig:3point1loope}+\ref{fig:3point1loopd}}  = \frac{g^2N}{4\pi^2}\left(2 H_j + \frac{1}{j+1}-1 \right)\,,
\end{equation}  
such that we recover the well-known one-loop contribution to the anomalous dimension of twist-two operators
\begin{align}\label{eqn:anomalous-dimension-Z-factor}
\gamma_j =  \left(\frac{g^2N}{4\pi^2}\right) 2H_j + \op(g^4)\,.
\end{align}
Using the tree-level three-point function and \eqref{equ:renormalised-op}, \eqref{eqn:mixing-matrix}, \eqref{eqn:renormalized-three-point-function} and \eqref{eqn:N(g^2)-for-d-dimensions} we find that the one-loop correction to the structure constant is
\begin{align}\label{eqn:result}
C_{\op\tilde{\op}j}^{(1)}/C_{\op\tilde{\op}j}^{(0)} = \frac{g^2 N}{8\pi^2} \left(5 H(j)^2 - 4 H(j) H(2j)- \sum_{r=1}^j \frac{1}{r^2} \right)\,,
\end{align}

\subsection{Normalisation Invariant Structure Constants}
For the structure constants appearing in the operator product expansion it is useful to take the normalisation of the two-point functions to be one.
The operators \eqref{eqn:definition-with-gegenbauer-polynomials} in terms of Gegenbauer polynomials are not normalised to one, but they have the perturbative expansion
\begin{equation}
 \langle \hat{\mathbb{O}}_j(x_1) \hat{\bar{\mathbb{O}}}_j(x_2) \rangle = \delta_{jk}  \left( C_j^{(0)} + g^2 C_j^{(1)} \right)2^{2j} \frac{  (\hat{x}_{12} )^{2j}}{(-x_{12}^2)^{2j+\theta}} + \op(g^4)\,,
\end{equation}
and one can explicitly calculate $C_j^{(0)}, C_j^{(1)}$. Using the Schwinger parametrisation of the propagator and some properties of the Gegenbauer polynomials as e.g. in \cite{Belitsky:2007jp} one finds that the operators \eqref{eqn:definition-with-gegenbauer-polynomials} have the tree-level normalisation
\begin{equation}\label{eqn:normalization-tree-level-two-point}
C_j^{(0)} = g^4 \delta^{aa} \frac{\Gamma(2j+1)}{2^{5} \pi^4 }\,.
\end{equation}

\begin{figure}[t]
\centering
\subfloat[~]{
\includegraphics[width=.2 \textwidth]{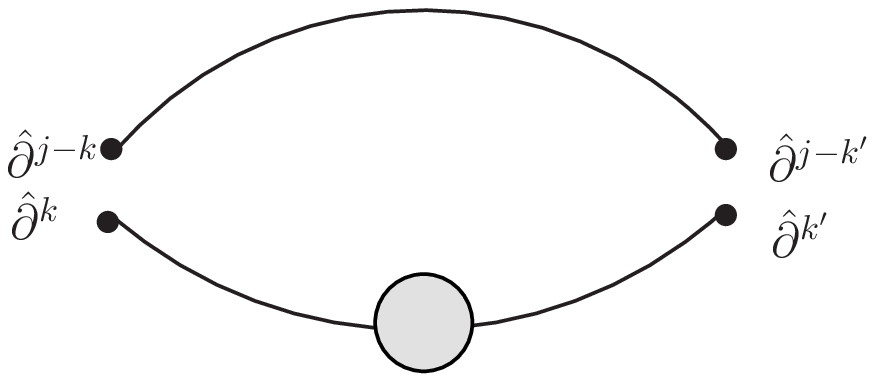}
\label{fig:twopointfunctionselfenergy}} 
~~~
\subfloat[~]{
\includegraphics[width=.2 \textwidth]{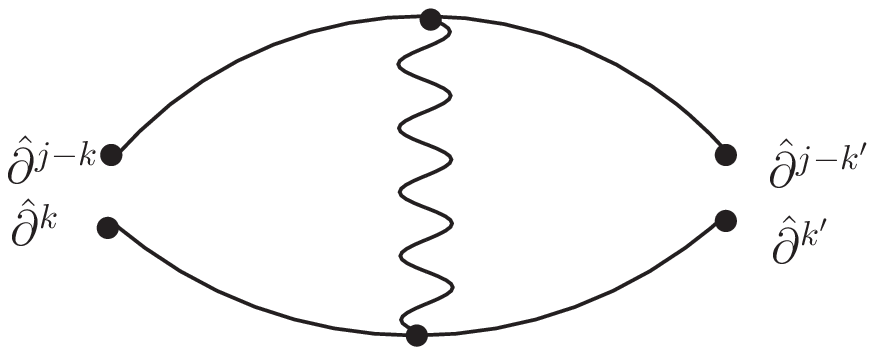}
\label{fig:twopointcourt}} \\
\subfloat[~]{
\includegraphics[width=.25 \textwidth]{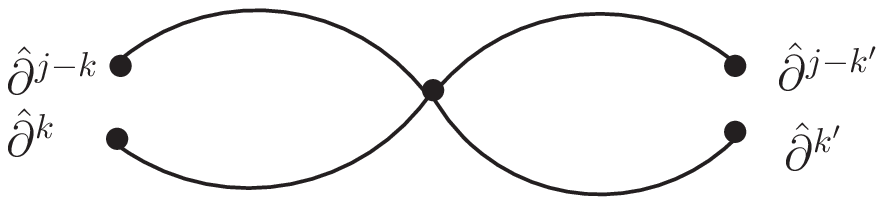}
\label{fig:twopointfourscalar}}
~~~
\subfloat[~]{
\includegraphics[width=.2 \textwidth]{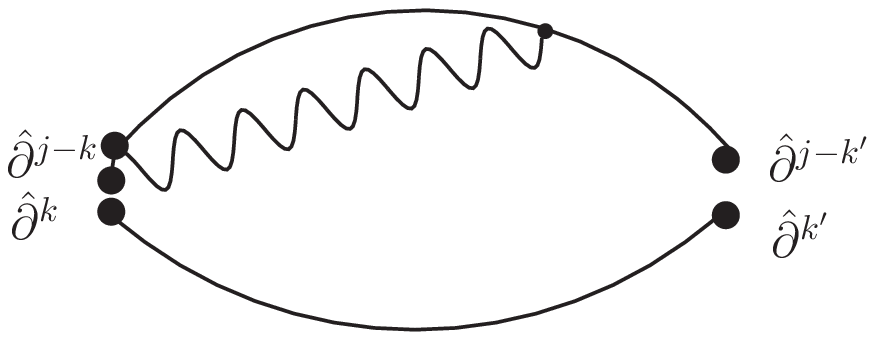}
\label{fig:twopointonecovderivative}}~~~
\subfloat[~]{
\includegraphics[width=.2 \textwidth]{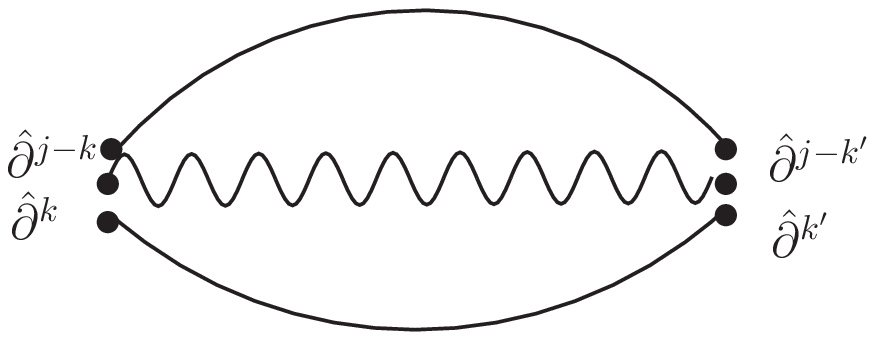}
\label{fig:twopointtwocovderivative}} \caption{One-loop corrections to the two-point function of $\langle \hat{\op}_j \hat{\op}_j \rangle$.}
\label{fig:two-point-function}
\end{figure}

At one-loop level the diagrams shown in figure \ref{fig:two-point-function} appear.
The calculation is well-known, technically similar to the three-point calculation and all appearing integrals can be found in the appendix \ref{sec:integrals}. We shall not reproduce the complete calculation here, but refer the reader to \cite{wiegandt:2012phdthesis} for more details on this and all other calculations. For the one-loop result of the normalisation we find
\begin{align}\label{eqn:result-norm}
C_j^{(1)}/C_j^{(0)} = \frac{g^2 N}{4\pi^2} \left(3H(j)^2 -2 H(j) H(2j) \right)\,.
\end{align}
The BPS operators \eqref{eqn:BPS-operators} have the normalisation $C=2^{-6}/\pi^4$ and do not get quantum corrections. We can define a normalisation invariant structure constant
$C_{\op\tilde{\op}j}^\prime$ which is related to  $C_{\op\tilde{\op}j}$ through 
\begin{align}\label{eqn:normalization invariant structure constant}
 C_{\op\tilde{\op}j}^\prime &= \frac{ C_{\op\tilde{\op}j}(g^2)}{\sqrt{C_j(g^2)}C}= \frac{C_{\op\tilde{\op}j}^{(0)}}{\sqrt{C_j^{(0)}}C} \left(1+ g^2\left( \frac{C_{\op\tilde{\op}j}^{(1)}}{C_{\op\tilde{\op}j}^{(0)}}-\frac{1}{2}\frac{C^{(1)}_j}{C^{(0)}_j} \right) \right) = C_{\op\tilde{\op}j}^{\prime (0)} + g^2 C_{\op\tilde{\op}j}^{\prime (1)}\,.
\end{align}
and corresponds to the structure constants, where all operators are normalised to one. Inserting \eqref{eqn:result-norm} and  \eqref{eqn:result} we thus find
\begin{equation}
C^\prime_{\op  \tilde{\op} j}(g^2) = {C^\prime}_{\op \tilde{\op} j}^{(0)} \left(1 +   \frac{g^2 N}{8 \pi^2}  \left(2H_j(H_j -  H_{2j})- H_{j,2}\right) + \op(g^4) \right)\,,
\end{equation} 
which is the result quoted in the introduction.

\section*{Acknowledgements}
We are grateful for many crucial discussions with Johannes Henn and Gregory Korchemsky. Furthermore, we would like to thank James Drummond for pointing out the extraction of the one-loop structure constants from the operator product expansion in \cite{Dolan:2004iy} to us.
 
The work of K.W. was partly supported by the German Science
Foundation (DFG) under the Collaborative Research Center (SFB) 676 "Particles,
Strings and the Early Universe" as well as (SFB) 647 "Space - Time - Matter. Analytic and Geometric Structures".
K.W. is grateful for hospitality at DESY Hamburg, where part of this work was carried out.

This work was supported by the Volkswagen foundation.

\appendix

\section{Notation and Conventions}\label{sec:conventions}
Throughout this paper we use the Minkowski-space metric with signature $(+ - - ...)$.
\subsection{$\sym$ Lagrangian}
We use the following action of $\sym$ theory
\begin{align} \label{eqn:N=4SYM-Lagrangian} 
S = \frac{1}{g^2}\int d^dx \,   \Big( &- \frac{1}{4} F_{\mu\nu}^aF^{a,\mu\nu} + \frac{1}{4} D_\mu \phi^{a,AB} D^\mu \bar{\phi}^{a}_{AB} \\ \nn
& 
- \frac{1}{16} f^{abc}f^{ade} \phi^{b,AB} \phi^{c,CD} \bar{\phi}^{d}_{AB} \bar{\phi}^{e}_{CD}  +   i \bar{\lambda}^a_{\dot{\alpha}A} (\sigma_\mu)^{\dot{\alpha}\beta} D^\mu \lambda_\beta^{a,A} 
\\ \nn
& - \frac{i}{\sqrt{2}} f^{abc} \lambda^{a,\alpha A} \bar{\phi}^{b}_{AB} \lambda_\alpha^{c,B} + \frac{i}{\sqrt{2}} f^{abc} \bar{\lambda}^{a}_{A,\dot{\alpha}} \phi^{b,AB} \bar{\lambda}^{c,\dot{\alpha}}_{B} \Big)\,,
\end{align}
where we have taken the trace of the matrix valued fields, e.g. $\bar{\phi}_{AB}=\bar{\phi}_{AB}^{a}T^a$ using the normalisation of the generators $T^a$ in the fundamental representation according to $\tr(T^a T^b) = \delta^{ab}/2$.

\subsection{Light-Cone Projection}\label{sec:light-cone-projection}
The calculations can be considerably simplified by projecting the indices to the light-cone using a light-like vector
\begin{equation}
\hat{\op}_j = \op_{\mu_1...\mu_j} z^{\mu_1}...z^{\mu_j} \,,\qquad z^2=0\,.
\end{equation}
The indices can be recovered by repeated application of a second order differential operator in $z^\mu$ 
\begin{equation}
\Delta_\mu = ((h-1) + z \cdot \partial)\partial_{z,\mu} - \frac{1}{2} z_\mu \partial_z \cdot \partial_z
\end{equation}
in the presence of the constraint $z^2=0$, see e.g.  \cite{Dobrev:1975ru}. In the following, we denote with a \emph{hat} the contraction with $z_\mu$, e.g.
\begin{equation}
\hat{x} = x_\mu z^\mu\,, \qquad \hat{\partial}_x =  z^\mu \partial_{x,\mu} =  z^\mu \frac{\partial}{\partial x^\mu}\,.
\end{equation}
The twist-two operator $\hat{\op}_j$ projected to the light-cone then reads
\begin{equation}\label{eqn:def-twist-two}
\hat{\op}_j (x) = \sum_{k=0}^j a_{jk}^{1/2}\,\tr\left( \hat{D}^k \phi^{12}(x) \hat{D}^{j-k} \phi^{12}(x)\right)\,,
\end{equation}
where $\hat{D}=D^\mu z_\mu$ is the light-cone projected covariant derivative and the numerical coefficients $a_{jk}^\nu$ are related to the so-called \emph{Gegenbauer polynomials} $C_j^\nu(x)$ such that
\begin{equation}\label{eqn:relation-gegenbauer-polynomials}
\sum_{k=0}^j a_{jk}^\nu\, x^k y^{j-k} = (x+y)^j C_j^{\nu} \left(\frac{x-y}{x+y}\right)
\end{equation}
and $\nu= h-3/2$, where $h=d/2$. 
Therefore, we can rewrite the operators in the bi-local form \cite{Ohrndorf:1981qv,Makeenko:1980bh}
\begin{align}\label{eqn:definition-with-gegenbauer-polynomials}
\hat{\op}_j^{\text{tree}} &= \left(\hat{\partial}_a+\hat{\partial}_b \right)^j C_j^{1/2} \left( \frac{\hat{\partial}_a-\hat{\partial}_b}{\hat{\partial}_a+\hat{\partial}_b} \right) \tr \left( \phi^{12}(x_a)  \phi^{12}(x_b) \right) \Big|_{x_a=x_b}\,.
\end{align}
Explicitly, the coefficients are given by $a_{jk}^{1/2} = (-1)^{k}{j \choose k }{j \choose k }$.  These operators have conformal two-point functions which are of the form
\begin{equation}\label{eqn:conformal two-point functions-tree}
 \langle \hat{\op}_j(x_1) \hat{\op}_k(x_2) \rangle = \delta_{jk} C_j 2^{2j} \frac{(  \hat{x}_{12} )^{2j}}{(-x_{12}^2)^{2j+\theta}}\qquad (j~ \text{even})\,,
\end{equation}
and the three-point functions read
\begin{equation}
 \langle \op(x_1)  \tilde{\op}(x_2) \hat{\op}_j(x_3) \rangle = C_{\op \tilde{\op} j} \frac{(  \hat{Y}_{12,3}) ^j}{|x_{12}|^{\Delta_1 + \Delta_2 - \theta} |x_{13}|^{\Delta_1 +  \theta- \Delta_2} |x_{23}|^{ \Delta_2 + \theta - \Delta_1 }}\,,
\end{equation}
where $|x_{ij}|=(-x_{ij}^2)^{1/2}$ and $\hat Y_{12;3} = Y^\mu(x_{13},x_{23}) z_\mu$.

\section{Details of the Calculation}

\subsection{Fourier Transformation and Bubble Integrals}
In Minkowski-space with signature $(+ - - ...)$ we have 
\begin{equation}\label{eqn:result-fourier-trafo-mink}
 \int \frac{d^dp}{(2\pi)^d} \frac{e^{-ip \cdot x}}{(-p^2-i\epsilon)^k} =  \frac{\Gamma(\frac{d}{2}-k)}{\Gamma(k)} \frac{1}{4^k \pi ^\frac{d}{2}} \frac{i}{(-x^2+i\epsilon)^{\frac{d}{2}-k}} \,.
\end{equation}
The two-point integral with momenta in the numerator is
\begin{align}\label{eqn:bubble-integral}
B_n(\alpha_1,\alpha_2)&= \int \frac{d^dk}{(2\pi)^d} \frac{\hat{k}^n}{(-k^2-i\epsilon)^{\alpha_1}(-(p+k)^2-i\epsilon)^{\alpha_2}} \\\nn
&= b_n(\alpha_1, \alpha_2) \frac{\hat{p}^n}{(-p^2-i\epsilon)^{\alpha_1+\alpha_2-\frac{d}{2}}}\,,
\end{align}
where
\begin{equation}\label{eqn:bna1a2}
 b_n(\alpha_1,\alpha_2) = i\frac{(-1)^n}{(4\pi)^{d/2}} \frac{\Gamma(\frac{d}{2}+n-\alpha_1)\Gamma(\frac{d}{2}-\alpha_2)}{\Gamma(d+n-\alpha_1-\alpha_2)} \frac{\Gamma(\alpha_1+\alpha_2-\frac{d}{2})}{\Gamma(\alpha_1)\Gamma(\alpha_2)}\,.
\end{equation}
Note, that analogous expression for the bubble integral and the Fourier transformation in $x$-space differ by minus signs, due to the different sign of $i \epsilon$ in the propagators.

\subsection{Normalization Factor of the $x_3$ Integration}
Integration of \eqref{eqn:three-point-structure} over $x_3$ yields
\begin{equation}
 \int d^dx_3 \langle \op \tilde{\op} \hat{\mathbb{O}}_j\rangle = N(g^2) \left(C_{\op\tilde{\op}j}^{(0)}+g^2 C_{\op\tilde{\op}j}^{(1)} \right)\frac{(\hat{x}_{12})^j}{(-x_{12}^2)^{j+d-3+\gamma_j(g^2)/2}} 
\end{equation}
where
\begin{equation}\label{eqn:N(g^2)-for-d-dimensions}N(g^2,d)= -i\frac{\Gamma(\theta-d/2+j) \Gamma((d-\theta)/2)^2 \Gamma \left( j+ (\theta -1)/2\right)}{\Gamma(d-\theta) \Gamma \left(j+\frac{\theta }{2}\right) \Gamma (j+\theta -1)} \frac{2^{\theta +2 j-2} }{\pi^{\frac{1}{2}-\frac{d}{2}} }\,.
 \end{equation}

\section{Integrals by using the IBP Method}\label{sec:integrals}

We define the following set of integrals\footnote{These integrals were considered before in \cite{Kazakov:1986mu}.}
\begin{equation}\label{eqn:basic-integral}
f_{mn}(a_1,a_2,a_3,a_4,a_5) = \int \frac{d^dk,l}{(2\pi)^{2d}} \frac{(\hat{k})^m (\hat{l})^n}{k^{2 a_1}(p+k)^{2a_2}(l-k)^{2a_3}l^{2 a_4}(p+l)^{2a_5}}\,,
\end{equation}
where $\hat{k}=z^\mu k_\mu$ is the contraction with a light-like vector $z^2=0$. The integral is symmetric under the simultaneous exchange  $(m,a_1,a_2) \leftrightarrow (n,a_4,a_5)$.
In particular in this work we need $f_{mn}(1,1,1,1,1)$ and $f_{j0}(2,1,1,1,1)$.  Since the momentum dependence is the same everywhere we define $c_{mn}$ by stripping off the equal coefficients 
\begin{equation}\label{eqn:definition-cmn}
 f_{mn}(a_1,a_2,a_3,a_4,a_5) = c_{mn}(a_1,a_2,a_3,a_4,a_5) \frac{(\hat{p})^{m+n}}{(-p^2)^{\sum_i a_i -d}} \,.
\end{equation}

\subsection{Bubble Integrals}\label{sec:app-reduced-integrals}
The integrals \eqref{eqn:basic-integral} with one argument set to zero are easy to solve and are the building blocks of $f_{mn}(1,1,1,1,1)$ after using the IBP method, see e.g. \cite{Smirnov:2006ry} for an introduction to the IBP technique. We also give more details on this technique in \cite{wiegandt:2012phdthesis}. We have
\begin{align}
c_{mn}(a_1,a_2,0,a_4,a_5) &= (-1)^{\sum_i a_i} b_m(a_1,a_2) b_n(a_4,a_5) 
\end{align}
and
\begin{equation}\nn
c_{mn}(a_1,a_2,a_3,a_4,0) = (-1)^n (-1)^{\sum_i a_i} b_n(a_4,a_3) b_{m+n}(a_1+a_3+a_4-\frac{d}{2},a_2)  \,.
\end{equation}
By the symmetry of the integral \eqref{eqn:basic-integral} $(m,a_1,a_2) \leftrightarrow (n,a_4,a_5)$ this also yields
\begin{equation}
c_{mn}(a_1,0,a_3,a_4,a_5) = c_{nm}(a_4,a_5,a_3,a_1,0)\,.
\end{equation}
Furthermore, we have
\begin{align} \nn
 c_{mn}(a_1,a_2,a_3,0,a_5) &= (-1)^{\sum_i a_i}  \quad  \sum_{j=0}^n \sum_{i=0}^{n-j} {n \choose j} {n -j \choose i} (-1)^n \\
& b_{n-j}(a_5,a_3) b_{m+n-i-j}(a_1,a_2+a_3+a_5-\frac{d}{2})
\end{align}
and $ c_{mn}(0,a_2,a_3,a_4,a_5)$ is related to this integral by the symmetry  $(m,a_1,a_2) \leftrightarrow (n,a_4,a_5)$.
We can perform the sum over $i$ and find
\begin{align}\label{eqn:fmna1a2a30a5}
 c_{mn}(a_1,a_2,a_3,0,a_5) &=   \frac{\Gamma(a_1-m)\Gamma(a_1+a_2+a_3+a_5-d)}{\Gamma(a_1)\Gamma(a_1+a_2+a_3+a_5-d-m)}  \\ \nn 
 & \sum_{j=0}^n  {n \choose j} (-1)^{m+n-j} b_{j}(a_5,a_3) b_{j}(a_2+a_3+a_5-\frac{d}{2},a_1-m)\,.
\end{align}
The sum can be performed, e.g. using Mathematica, in terms of  generalised hypergeometric functions.

\subsection{Integrals $f_{mn}(1,1,1,1,1)$}\label{sec:Integrals for diagrams b}
First we would like to solve the integral  $f_{mn}(1,1,1,1,1)$, given through the definition
\begin{equation}
 f_{mn}(a_1,a_2,a_3,a_4,a_5) = \int \frac{d^dk,l}{(2\pi)^{2d}} \underbrace{\frac{(\hat{k})^m (\hat{l})^n}{k^{2 a_1} (p+k)^{2 a_2} (k-l)^{2 a_3} l^{2 a_4} (p+l)^{2 a_5}}}_{\hat{i}_{mn}(....)}\,.
\end{equation}
The integral is symmetric under simultaneous exchange of $\{m,a_1,a_2\} \leftrightarrow \{n,a_4,a_5\}$ and can be solved by using the IBP method.
\begin{equation}\label{eqn:IBP-relation}
(d+m- 4) f_{mn} =\text{bubbles}(m,n) + m \, f_{m-1,n+1}\,,
\end{equation}
where we abbreviate the known integrals as
\begin{align}\label{eqn:bubbles}
\text{bubbles}(m,n) &= f_{mn}(2,1,0,1,1) - f_{mn}(2,1,1,0,1)\\ \nonumber
&+  f_{mn}(1,2,0,1,1)-  f_{mn}(1,2,1,1,0)\,.
\end{align}
We can actually solve for all integrals by writing down \eqref{eqn:IBP-relation} for $m=n+1$ and $n=m-1$, i.e.
\begin{align}\label{eqn:recursion2}
(d+(n+1)- 4) f_{n+1,m-1} &= \text{bubbles}(n+1,m-1)+ (n+1) \, f_{nm}
\end{align}
and using $f_{mn}(1,1,1,1,1)=f_{nm}(1,1,1,1,1)$. Solving \eqref{eqn:recursion2} for $f_{n+1,m-1}$  and inserting it into \eqref{eqn:IBP-relation} we get $f_{nm}$ in terms of known integrals:
\begin{equation}\label{eqn:solution-fnm}
f_{mn}(1,1,1,1,1) = \frac{(n+d-3)\text{bubbles}(m,n)+m \text{bubbles}(n+1,m-1)}{(d-4)(d-3+m+n)}\,.
\end{equation}
Note, that the coefficient on the left-hand side is of order $\epsilon$ in  $d=4-2\epsilon$.
All appearing integrals were solved in section \ref{sec:app-reduced-integrals}.

\subsection{Integrals $f_{mn}(2,1,1,1,1)$}\label{eqn:integral-f_{j,0}(2,1,1,1,1)}

For the three-point function calculation we need the integral
\begin{equation}
f_{j,0}(2,1,1,1,1) = \int \frac{d^dk,l}{(2\pi)^{2d}} \frac{(\hat{k})^j}{k^4 (p+k)^2(l-k)^2 l^2(p+l)^2}\,.
\end{equation}
Using the IBP method, just as in section \ref{sec:Integrals for diagrams b} for $f_{mn}(1,1,1,1,1)$, we find
\begin{equation}\label{eqn:recursion-fj0(2,1,1,1,1)}
(d-5+m) f_{m,n}(2,1,1,1,1) = m f_{m-1,n+1}(2,1,1,1,1) + \text{bubbles}_2(m,n)\,,
\end{equation}
where
\begin{align}
\text{bubbles}_2(m,n) &=  2 f_{mn} (3,1,0,1,1) - 2 f_{mn} (3,1,1,0,1) \\ \nn
&\phantom{= } +f_{mn} (2,2,0,1,1)-f_{mn} (2,2,1,1,0)
\end{align}
and all these integrals are solved in the following section. Thus we can obtain $f_{mn}$ recursively from \eqref{eqn:recursion-fj0(2,1,1,1,1)}.

\end{document}